\documentclass[pra,showpacs,twocolumn,superscriptaddress]{revtex4}

\def\temp{1.35}%
\let\tempp=\relax
\expandafter\ifx\csname psboxversion\endcsname\relax
\else
    \ifdim\temp cm>\psboxversion cm
    \else
      \let\temp=\psboxversion
      \let\tempp= 
    \fi
\fi
\tempp
\let\psboxversion=\temp
\catcode`\@=11
\def\psfortextures{%     For TeXtures on the Macintosh
\def\PSspeci@l##1##2{%
\special{illustration ##1\space scaled ##2}%
}}%
\def\psfordvitops{%      For the DVItoPS converter on IBM mainframes
\def\PSspeci@l##1##2{%
\special{dvitops: import ##1\space \the\drawingwd \the\drawinght}%
}}%
\def\psfordvips{%      For DVIPS converter on VAX, UNIX and PC's
\def\PSspeci@l##1##2{%
\d@my=0.1bp \d@mx=\drawingwd \divide\d@mx by\d@my% BUG! for large \drawingwd
\includegraphics{##1\space}}}%
\def\psforoztex{%        For the OzTeX shareware on the Macintosh
\def\PSspeci@l##1##2{%
\special{##1 \space
      ##2 1000 div dup scale
      \number-\psllx\space\space \number-\pslly\space\space translate
}}}%
\def\psfordvitps{%       From the UNIX TeXPS package, vers.>3.12
\def\dvitpsLiter@ldim##1{\dimen0=##1\relax
\special{dvitps: Literal "\number\dimen0\space"}}%
\def\PSspeci@l##1##2{%
\at(0bp;\drawinght){%
\special{dvitps: Include0 "psfig.psr"}% contains def of "startTexFig"
\dvitpsLiter@ldim{\drawingwd}%
\dvitpsLiter@ldim{\drawinght}%
\dvitpsLiter@ldim{\psllx bp}%
\dvitpsLiter@ldim{\pslly bp}%
\dvitpsLiter@ldim{\psurx bp}%
\dvitpsLiter@ldim{\psury bp}%
\special{dvitps: Literal "startTexFig"}%
\special{dvitps: Include1 "##1"}%
\special{dvitps: Literal "endTexFig"}%
}}}%
\def\psfordvialw{%   Try for dvialw, a UNIX public domain
\def\PSspeci@l##1##2{
\special{language "PostScript",
position = "bottom left",
literal "  \psllx\space \pslly\space translate
  ##2 1000 div dup scale
  -\psllx\space -\pslly\space translate",
include "##1"}
}}%
\def\psforptips{%   For MS-DOS; LUOMA@brandeis.bitnet
\def\PSspeci@l##1##2{{
\d@mx=\psurx bp
\advance \d@mx by -\psllx bp
\divide \d@mx by 1000\multiply\d@mx by \xscale
\incm{\d@mx}
\let\tmpx\dimincm
\d@my=\psury bp
\advance \d@my by -\pslly bp
\divide \d@my by 1000\multiply\d@my by \xscale
\incm{\d@my}
\let\tmpy\dimincm
\d@mx=-\psllx bp
\divide \d@mx by 1000\multiply\d@mx by \xscale
\d@my=-\pslly bp
\divide \d@my by 1000\multiply\d@my by \xscale
\at(\d@mx;\d@my){\special{ps:##1 x=\tmpx cm, y=\tmpy cm}}
}}}%
\def\psonlyboxes{%     Draft-like behaviour if none of the others works
\def\PSspeci@l##1##2{%
\at(0cm;0cm){\boxit{\vbox to\drawinght
  {\vss\hbox to\drawingwd{\at(0cm;0cm){\hbox{({\tt##1})}}\hss}}}}
}}%
\def\psloc@lerr#1{%
\let\savedPSspeci@l=\PSspeci@l%
\def\PSspeci@l##1##2{%
\at(0cm;0cm){\boxit{\vbox to\drawinght
  {\vss\hbox to\drawingwd{\at(0cm;0cm){\hbox{({\tt##1}) #1}}\hss}}}}
\let\PSspeci@l=\savedPSspeci@l% restore normal output for other figs!
}}%
\newread\pst@mpin
\newdimen\drawinght\newdimen\drawingwd
\newdimen\psxoffset\newdimen\psyoffset
\newbox\drawingBox
\newcount\xscale \newcount\yscale \newdimen\pscm\pscm=1cm
\newdimen\d@mx \newdimen\d@my
\newdimen\pswdincr \newdimen\pshtincr
\let\ps@nnotation=\relax
{\catcode`\|=0 |catcode`|\=12 |catcode`|%=12 |catcode`~=12
|catcode`#=12 |catcode`*=14
|xdef|backslashother{\}*
|xdef|percentother{%}*
|xdef|tildeother{~}*
|xdef|sharpother{#}*
}%
\def\R@moveMeaningHeader#1:->{}%
\def\uncatcode#1{%
\edef#1{\expandafter\R@moveMeaningHeader\meaning#1}}%
\def\execute#1{#1}% NOT stupid: cs in #1 are then identified BEFORE execution
\def\psm@keother#1{\catcode`#112\relax}% borrowed from latex
\def\executeinspecs#1{%
\execute{\begingroup\let\do\psm@keother\dospecials\catcode`\^^M=9#1\endgroup}}%
\def\@mpty{}%
\def\matchexpin#1#2{
  \fi%
  \edef\tmpb{{#2}}%
  \expandafter\makem@tchtmp\tmpb%
  \edef\tmpa{#1}\edef\tmpb{#2}%
  \expandafter\expandafter\expandafter\m@tchtmp\expandafter\tmpa\tmpb\endm@tch%
  \if\match%
}%
\def\matchin#1#2{%
  \fi%
  \makem@tchtmp{#2}%
  \m@tchtmp#1#2\endm@tch%
  \if\match%
}%
\def\makem@tchtmp#1{\def\m@tchtmp##1#1##2\endm@tch{%
  \def\tmpa{##1}\def\tmpb{##2}\let\m@tchtmp=\relax%
  \ifx\tmpb\@mpty\def\match{YN}%
  \else\def\match{YY}\fi%
}}%
\def\incm#1{{\psxoffset=1cm\d@my=#1
 \d@mx=\d@my
  \divide\d@mx by \psxoffset
  \xdef\dimincm{\number\d@mx.}
  \advance\d@my by -\number\d@mx cm
  \multiply\d@my by 100
 \d@mx=\d@my
  \divide\d@mx by \psxoffset
  \edef\dimincm{\dimincm\number\d@mx}
  \advance\d@my by -\number\d@mx cm
  \multiply\d@my by 100
 \d@mx=\d@my
  \divide\d@mx by \psxoffset
  \xdef\dimincm{\dimincm\number\d@mx}
}}%
\newif\ifNotB@undingBox
\newhelp\PShelp{Proceed: you'll have a 5cm square blank box instead of
your graphics.}%
\def\s@tsize#1 #2 #3 #4\@ndsize{
  \def\psllx{#1}\def\pslly{#2}%
  \def\psurx{#3}\def\psury{#4}%  needed by a crazyness of dvips!
  \ifx\psurx\@mpty\NotB@undingBoxtrue% this is not a valid one!
  \else
    \drawinght=#4bp\advance\drawinght by-#2bp
    \drawingwd=#3bp\advance\drawingwd by-#1bp
  \fi
  }%
\def\sc@nBBline#1:#2\@ndBBline{\edef\p@rameter{#1}\edef\v@lue{#2}}%
\def\g@bblefirstblank#1#2:{\ifx#1 \else#1\fi#2}%
{\catcode`\%=12
\xdef\B@undingBox{%%BoundingBox}}%
\def\ReadPSize#1{
 \readfilename#1\relax
 \let\PSfilename=\lastreadfilename
 \openin\pst@mpin=#1\relax
 \ifeof\pst@mpin \errhelp=\PShelp
   \errmessage{I haven't found your postscript file (\PSfilename)}%
   \psloc@lerr{was not found}%
   \s@tsize 0 0 142 142\@ndsize
   \closein\pst@mpin
 \else
   \if\matchexpin{\GlobalInputList}{, \lastreadfilename}%
   \else\xdef\GlobalInputList{\GlobalInputList, \lastreadfilename}%
     \immediate\write\psbj@inaux{\lastreadfilename,}%
   \fi%
   \loop
     \executeinspecs{\catcode`\ =10\global\read\pst@mpin to\n@xtline}%
     \ifeof\pst@mpin
       \errhelp=\PShelp
       \errmessage{(\PSfilename) is not an Encapsulated PostScript File:
           I could not find any \B@undingBox: line.}%
       \edef\v@lue{0 0 142 142:}%
       \psloc@lerr{is not an EPSFile}%
       \NotB@undingBoxfalse
     \else
       \expandafter\sc@nBBline\n@xtline:\@ndBBline
       \ifx\p@rameter\B@undingBox\NotB@undingBoxfalse
         \edef\t@mp{%
           \expandafter\g@bblefirstblank\v@lue\space\space\space}%
         \expandafter\s@tsize\t@mp\@ndsize
       \else\NotB@undingBoxtrue
       \fi
     \fi
   \ifNotB@undingBox\repeat
   \closein\pst@mpin
 \fi
\message{#1}%
}%
\def\psboxto(#1;#2)#3{\vbox{%
   \ReadPSize{#3}%
   \advance\pswdincr by \drawingwd
   \advance\pshtincr by \drawinght
   \divide\pswdincr by 1000
   \divide\pshtincr by 1000
   \d@mx=#1
   \ifdim\d@mx=0pt\xscale=1000
         \else \xscale=\d@mx \divide \xscale by \pswdincr\fi
   \d@my=#2
   \ifdim\d@my=0pt\yscale=1000
         \else \yscale=\d@my \divide \yscale by \pshtincr\fi
   \ifnum\yscale=1000
         \else\ifnum\xscale=1000\xscale=\yscale
                    \else\ifnum\yscale<\xscale\xscale=\yscale\fi
              \fi
   \fi
   \divide\drawingwd by1000 \multiply\drawingwd by\xscale
   \divide\drawinght by1000 \multiply\drawinght by\xscale
   \divide\psxoffset by1000 \multiply\psxoffset by\xscale
   \divide\psyoffset by1000 \multiply\psyoffset by\xscale
   \global\divide\pscm by 1000
   \global\multiply\pscm by\xscale
   \multiply\pswdincr by\xscale \multiply\pshtincr by\xscale
   \ifdim\d@mx=0pt\d@mx=\pswdincr\fi
   \ifdim\d@my=0pt\d@my=\pshtincr\fi
   \message{scaled \the\xscale}%
 \hbox to\d@mx{\hss\vbox to\d@my{\vss
   \global\setbox\drawingBox=\hbox to 0pt{\kern\psxoffset\vbox to 0pt{%
      \kern-\psyoffset
      \PSspeci@l{\PSfilename}{\the\xscale}%
      \vss}\hss\ps@nnotation}%
   \global\wd\drawingBox=\the\pswdincr
   \global\ht\drawingBox=\the\pshtincr
   \global\drawingwd=\pswdincr
   \global\drawinght=\pshtincr
   \baselineskip=0pt
   \copy\drawingBox
 \vss}\hss}%
  \global\psxoffset=0pt
  \global\psyoffset=0pt
  \global\pswdincr=0pt
  \global\pshtincr=0pt % These are local to one figure
  \global\pscm=1cm %should not be necessary
}}%
\def\psboxscaled#1#2{\vbox{%
  \ReadPSize{#2}%
  \xscale=#1
  \message{scaled \the\xscale}%
  \divide\pswdincr by 1000 \multiply\pswdincr by \xscale
  \divide\pshtincr by 1000 \multiply\pshtincr by \xscale
  \divide\psxoffset by1000 \multiply\psxoffset by\xscale
  \divide\psyoffset by1000 \multiply\psyoffset by\xscale
  \divide\drawingwd by1000 \multiply\drawingwd by\xscale
  \divide\drawinght by1000 \multiply\drawinght by\xscale
  \global\divide\pscm by 1000
  \global\multiply\pscm by\xscale
  \global\setbox\drawingBox=\hbox to 0pt{\kern\psxoffset\vbox to 0pt{%
     \kern-\psyoffset
     \PSspeci@l{\PSfilename}{\the\xscale}%
     \vss}\hss\ps@nnotation}%
  \advance\pswdincr by \drawingwd
  \advance\pshtincr by \drawinght
  \global\wd\drawingBox=\the\pswdincr
  \global\ht\drawingBox=\the\pshtincr
  \global\drawingwd=\pswdincr
  \global\drawinght=\pshtincr
  \baselineskip=0pt
  \copy\drawingBox
  \global\psxoffset=0pt
  \global\psyoffset=0pt
  \global\pswdincr=0pt
  \global\pshtincr=0pt % These are local to one figure
  \global\pscm=1cm
}}%
\def\psbox#1{\psboxscaled{1000}{#1}}%
\newif\ifn@teof\n@teoftrue
\newif\ifc@ntrolline
\newif\ifmatch
\newread\j@insplitin
\newwrite\j@insplitout
\newwrite\psbj@inaux
\immediate\openout\psbj@inaux=psbjoin.aux
\immediate\write\psbj@inaux{\string\joinfiles}%
\immediate\write\psbj@inaux{\jobname,}%
\def\toother#1{\ifcat\relax#1\else\expandafter%
  \toother@ux\meaning#1\endtoother@ux\fi}%
\def\toother@ux#1 #2#3\endtoother@ux{\def\tmp{#3}%
  \ifx\tmp\@mpty\def\tmp{#2}\let\next=\relax%
  \else\def\next{\toother@ux#2#3\endtoother@ux}\fi%
\next}%
\let\readfilenamehook=\relax
\def\re@d{\expandafter\re@daux}% spares typing 10 \expandafter's...
\def\re@daux{\futurelet\nextchar\stopre@dtest}%
\def\re@dnext{\xdef\lastreadfilename{\lastreadfilename\nextchar}%
  \afterassignment\re@d\let\nextchar}%
\def\stopre@d{\egroup\readfilenamehook}%
\def\stopre@dtest{%
  \ifcat\nextchar\relax\let\nextread\stopre@d
  \else
    \ifcat\nextchar\space\def\nextread{%
      \afterassignment\stopre@d\chardef\nextchar=`}%
    \else\let\nextread=\re@dnext
      \toother\nextchar
      \edef\nextchar{\tmp}%
    \fi
  \fi\nextread}%
\def\readfilename{\bgroup%
  \let\\=\backslashother \let\%=\percentother \let\~=\tildeother
  \let\#=\sharpother \xdef\lastreadfilename{}%
  \re@d}%
\xdef\GlobalInputList{\jobname}%
\def\psnewinput{%
  \def\readfilenamehook{% each entry in \GlobalInputList should be unique
    \if\matchexpin{\GlobalInputList}{, \lastreadfilename}%
    \else\xdef\GlobalInputList{\GlobalInputList, \lastreadfilename}%
      \immediate\write\psbj@inaux{\lastreadfilename,}%
    \fi%
    \let\readfilenamehook=\relax%
    \ps@ldinput\lastreadfilename\relax%
  }\readfilename%
}%
\expandafter\ifx\csname @@input\endcsname\relax    % then Plain
  \immediate\let\ps@ldinput=\input\def\input{\psnewinput}%
\else
  \immediate\let\ps@ldinput=\@@input
  \def\@@input{\psnewinput}%
\fi%
\def\nowarnopenout{%
 \def\warnopenout##1##2{%
   \readfilename##2\relax
   \message{\lastreadfilename}%
   \immediate\openout##1=\lastreadfilename\relax}}%
\def\warnopenout#1#2{%
 \readfilename#2\relax
 \def\t@mp{TrashMe,psbjoin.aux,psbjoint.tex,}\uncatcode\t@mp
 \if\matchexpin{\t@mp}{\lastreadfilename,}%
 \else
   \immediate\openin\pst@mpin=\lastreadfilename\relax
   \ifeof\pst@mpin
     \else
     \edef\tmp{{If the content of this file is precious to you, this
is your last chance to abort (ie press x or e) and rename it before
retexing (\jobname). If you're sure there's no file
(\lastreadfilename) in the directory of (\jobname), then go on: I'm
simply worried because you have another (\lastreadfilename) in some
directory I'm looking in for inputs...}}%
     \errhelp=\tmp
     \errmessage{I may be about to replace your file named \lastreadfilename}%
   \fi
   \immediate\closein\pst@mpin
 \fi
 \message{\lastreadfilename}%
 \immediate\openout#1=\lastreadfilename\relax}%
{\catcode`\%=12\catcode`\*=14
\gdef\splitfile#1{*
 \readfilename#1\relax
 \immediate\openin\j@insplitin=\lastreadfilename\relax
 \ifeof\j@insplitin
   \message{! I couldn't find and split \lastreadfilename!}*
 \else
   \immediate\openout\j@insplitout=TrashMe
   \message{< Splitting \lastreadfilename\space into}*
   \loop
     \ifeof\j@insplitin
       \immediate\closein\j@insplitin\n@teoffalse
     \else
       \n@teoftrue
       \executeinspecs{\global\read\j@insplitin to\spl@tinline\expandafter
         \ch@ckbeginnewfile\spl@tinline%Beginning-Of-File-Named:%\endcheck}*
       \ifc@ntrolline
       \else
         \toks0=\expandafter{\spl@tinline}*
         \immediate\write\j@insplitout{\the\toks0}*
       \fi
     \fi
   \ifn@teof\repeat
   \immediate\closeout\j@insplitout
 \fi\message{>}*
}*
\gdef\ch@ckbeginnewfile#1%Beginning-Of-File-Named:#2%#3\endcheck{*
 \def\t@mp{#1}*
 \ifx\@mpty\t@mp
   \def\t@mp{#3}*
   \ifx\@mpty\t@mp
     \global\c@ntrollinefalse
   \else
     \immediate\closeout\j@insplitout
     \warnopenout\j@insplitout{#2}*
     \global\c@ntrollinetrue
   \fi
 \else
   \global\c@ntrollinefalse
 \fi}*
\gdef\joinfiles#1\into#2{*
 \message{< Joining following files into}*
 \warnopenout\j@insplitout{#2}*
 \message{:}*
 {*
 \edef\w@##1{\immediate\write\j@insplitout{##1}}*
\w@{% This collection of files was produced with CERN psbox package}*
\w@{% To decompose and tex it:}*
\w@{%-save this with a filename CONTAINING ONLY LETTERS and a .TEX}*
\w@{% extension (say, JOINTFIL.TEX), in some empty directory;}*
\w@{%-make sure you can \string\input\space psbox.tex (version>=1.3);}*
\w@{%  (else ftp cs.nyu.edu(=128.122.140.24):pub/TeX/psbox/, then get}*
\w@{%  and tex the file psboxall.tex; more info in psbREAD.ME)}*
\w@{%-tex JOINTFIL.TEX using Plain, or LaTeX, or whatever is needed by}*
\w@{%  the first file in the joining (after splitting JOINTFIL.TEX into}*
\w@{%  it's constituents, TeX will try to process it as it stands).}*
\w@{\string\input\space psbox.tex}*
\w@{\string\splitfile{\string\jobname}}*
\w@{\string\let\string\autojoin=\string\relax}*
}*
 \expandafter\tre@tfilelist#1, \endtre@t
 \immediate\closeout\j@insplitout
 \message{>}*
}*
\gdef\tre@tfilelist#1, #2\endtre@t{*
 \readfilename#1\relax
 \ifx\@mpty\lastreadfilename
 \else
   \immediate\openin\j@insplitin=\lastreadfilename\relax
   \ifeof\j@insplitin
     \errmessage{I couldn't find file \lastreadfilename}*
   \else
     \message{\lastreadfilename}*
     \immediate\write\j@insplitout{%Beginning-Of-File-Named:\lastreadfilename}*
     \executeinspecs{\global\read\j@insplitin to\oldj@ininline}*
     \loop
       \ifeof\j@insplitin\immediate\closein\j@insplitin\n@teoffalse
       \else\n@teoftrue
         \executeinspecs{\global\read\j@insplitin to\j@ininline}*
         \toks0=\expandafter{\oldj@ininline}*
         \let\oldj@ininline=\j@ininline
         \immediate\write\j@insplitout{\the\toks0}*
       \fi
     \ifn@teof
     \repeat
   \immediate\closein\j@insplitin
   \fi
   \tre@tfilelist#2, \endtre@t
 \fi}*
}%
\def\autojoin{%
 \immediate\write\psbj@inaux{\string\into{psbjoint.tex}}%
 \immediate\closeout\psbj@inaux
 \expandafter\joinfiles\GlobalInputList\into{psbjoint.tex}%
}%
\def\centinsert#1{\midinsert\line{\hss#1\hss}\endinsert}%
\def\psannotate#1#2{\vbox{%
  \def\ps@nnotation{#2\global\let\ps@nnotation=\relax}#1}}%
\def\pscaption#1#2{\vbox{%
   \setbox\drawingBox=#1
   \copy\drawingBox
   \vskip\baselineskip
   \vbox{\hsize=\wd\drawingBox\setbox0=\hbox{#2}%
     \ifdim\wd0>\hsize
       \noindent\unhbox0\tolerance=5000
    \else\centerline{\box0}%
    \fi
}}}%
\def\at(#1;#2)#3{\setbox0=\hbox{#3}\ht0=0pt\dp0=0pt
  \rlap{\kern#1\vbox to0pt{\kern-#2\box0\vss}}}%
\newdimen\gridht \newdimen\gridwd
\def\gridfill(#1;#2){%
  \setbox0=\hbox to 1\pscm
  {\vrule height1\pscm width.4pt\leaders\hrule\hfill}%
  \gridht=#1
  \divide\gridht by \ht0
  \multiply\gridht by \ht0
  \gridwd=#2
  \divide\gridwd by \wd0
  \multiply\gridwd by \wd0
  \advance \gridwd by \wd0
  \vbox to \gridht{\leaders\hbox to\gridwd{\leaders\box0\hfill}\vfill}}%
\def\fillinggrid{\at(0cm;0cm){\vbox{%
  \gridfill(\drawinght;\drawingwd)}}}%
\def\textleftof#1:{%
  \setbox1=#1
  \setbox0=\vbox\bgroup
    \advance\hsize by -\wd1 \advance\hsize by -2em}%
\def\textrightof#1:{%
  \setbox0=#1
  \setbox1=\vbox\bgroup
    \advance\hsize by -\wd0 \advance\hsize by -2em}%
\def\endtext{%
  \egroup
  \hbox to \hsize{\valign{\vfil##\vfil\cr%
\box0\cr%
\noalign{\hss}\box1\cr}}}%
\def\frameit#1#2#3{\hbox{\vrule width#1\vbox{%
  \hrule height#1\vskip#2\hbox{\hskip#2\vbox{#3}\hskip#2}%
        \vskip#2\hrule height#1}\vrule width#1}}%
\def\boxit#1{\frameit{0.4pt}{0pt}{#1}}%
\catcode`\@=12 % cs containing @ are unreachable
\psfordvips   %     For DVIPS converter on VAX and UNIX
\usepackage{graphics}
\usepackage{multirow}
\usepackage[usenames]{color}
\usepackage{nicefrac}
\usepackage{amsmath}

\newcommand {\mb}[1]{\mbox{\boldmath{${#1}$}}}

\begin{document}

\title{Approximate mean--field equations of motion for quasi--2D 
Bose--Einstein condensate systems}
\author{Mark Edwards}
\affiliation{Department of Physics, Georgia Southern University,
Statesboro, GA 30460--8031 USA}
\author{Michael Krygier}
\affiliation{Department of Physics, Georgia Southern University,
Statesboro, GA 30460--8031 USA}
\author{Hadayat Seddiqi}
\affiliation{Department of Physics, Georgia Southern University,
Statesboro, GA 30460--8031 USA}
\author{Brandon Benton}
\affiliation{Department of Physics, Georgia Southern University,
Statesboro, GA 30460--8031 USA}
\author{Charles W.\ Clark}
\affiliation{Joint Quantum Insitute, National Institute of Standards 
and Technology and the University of Maryland, Gaithersburg, MD 20899, USA}

\date{\today}

\begin{abstract}
We present a method for approximating the solution of the three--%
dimensional, time--dependent Gross--Pitaevskii equation (GPE) for 
Bose--Einstein condensate systems where the confinement in one dimension 
is much tighter than in the other two.  This method employs a hybrid
Lagrangian variational technique whose trial wave function is the 
product of a completely unspecified function of the coordinates in the
plane of weak confinement and a gaussian in the strongly confined 
direction having a time--dependent width and quadratic phase.  The 
hybrid Lagrangian variational method produces equations of motion that
consist of (1) a two--dimensional, effective GPE whose nonlinear coef%
ficient contains the width of the gaussian and (2) an equation of motion
for the width that depends on the integral of the fourth power of the
solution of the 2D effective GPE.  We apply this method to the dynamics
of Bose--Einstein condensates confined in ring--shaped potentials and
compare the approximate solution to the numerical solution of the full 
3D GPE.
\end{abstract}


\maketitle

\section{Introduction}
\label{intro}

Recent advances in laser--control technology have enabled the laboratory
realization of Bose--Einstein condensate (BEC) systems subjected to 
all--optical potentials which provide strong confinement in a horizontal 
plane and an arbitrary potential within this plane.  These potentials
can be produced by a combination of a horizontal light sheet combined with 
a rapidly moving red-- or blue--detuned vertical laser that ``paints'' 
an arbitrary time--averaged optical dipole potential in the horizontal 
plane~\cite{1367-2630-11-4-043030}.  Horizontal light sheets can also
be combined with vertically propagating beams in specialized laser modes,
such as Laguerre--Gauss modes, to produce other types of novel potentials%
~\cite{PhysRevLett.106.130401}.  In addition to providing strong vertical 
confinement and counteracting the effect of gravity, the light sheet 
provides stabilization against dynamic excitations of the condensate%
~\cite{PhysRevLett.84.810} as well as thermal phase fluctuations~\cite
{PhysRevLett.87.160406}.

The ability to create and probe quasi--2D BECs in arbitray 2D potentials
is motivated by several areas of current ultra--cold atom research.  For 
example, condensates in toroidal and ring lattices can be studied.  Stable 
states of multiple vortices and persistent currents can be created and studied 
by stirring the condensate~\cite{Brand_Reinhardt_2001,PhysRevLett.99.260401,
PhysRevLett.106.130401}.  There are proposals for creating ring lattices 
and for studying non--equilibrium phase transitions within this geometry
\cite{PhysRevLett.95.063201,PhysRevLett.101.115701}.  Toroidal geometries
are well--suited for studying topological defects that may appear during
a rapid cooling process that produces a condensate~\cite{PhysRevLett.83.1707,
Anderson_Nature_2008}.  There is some indication that stirring within a
ring--lattice geometry can produce a coherent superposition of states with
different circulation which can lead to a reduction in the threshold of
the Mott--Insulator phase transition~\cite{PhysRevA.75.063616}.  These 
systems also offer an excellent finite--sized testbed for systems of 
ultra--cold atoms that mimic condensed--matter systems~\cite{lewenstein_review}.

Quasi--2D BECs may also provide a convenient platform for studying systems 
of ultra--cold neutral atoms that are analogs of electronic materials, devices, 
and circuits~\cite{PhysRevA.75.023615}.  Such systems are called ``atomtronic'' 
because strongly interacting Bose gases in a lattice potential are analogous 
to ``electronic'' systems of electrons moving in the periodic lattice potential 
of a crystalline solid.  The ability to produce arbitrary potentials in 
the plane of the quasi--2D condensate may enable the controlled study of 
novel atomtronic systems.  In particular it may be possible to produce 
circuit--like potentials within the plane.

The behavior of many of the above--mentioned ultra--cold bosonic systems 
can be described using mean--field theory.  In this case, the governing
equation is the time--dependent Gross--Pitaevskii equation (TDGPE)~\cite
{gross_gp_paper,pita_gp_paper}.  This is a partial differential equation 
in three space variables and one time variable whose solution represents 
the wave function of the single--particle orbital that all of the condensate 
atoms occupy.  Experiments conducted on these systems typically involve
releasing the condensate for imaging.  In this case solution of the 3D 
TDGPE becomes a challenging numerical problem due to the volume that must
be accounted for in simulating the experiment.  

In this paper, we present a variational approximation to the solution of 
the TDGPE for these quasi--2D systems which produces equations of motion 
whose numerical solution can be obtained 100 to 1000 times faster than 
solving the full 3D TDGPE. This approximation is based on a variant of the 
standard Lagrangian Variational Method (LVM)~\cite{PhysRevA.56.1424} in 
which some of the variational parameters are functions of the space 
coordinates and time while others are only functions of time.  The work 
presented here applies this ``hybrid'' version of the LVM to a quasi--2D 
systems of bosonic atoms.  The hybrid LVM was previously applied to a 
quasi--1D system where only one dimension was weakly confined compared 
to the two other dimensions~\cite{1d_hlvm_paper}.

This paper is organized as follows.  In Section \ref{eom} we describe the
LVM and its hybrid form and derive the approximate equations of motion.
We also derive the equations that provide the proper variational stationary
solution in which a condensate is trapped in a confining potential.  
Section \ref{2d3d_compare} presents a comparison of the solution of the
hybrid LVM equations of motion with the numerical solution of the 3D TDGPE
for a BEC confined in a ring--shaped potential.  The parameters for this
example system were taken from an actual experiment.  Section \ref{summary}
presents a summary of the work.

\section{Two--dimensional hybrid LVM equations of motion}
\label{eom}

The condensate wave function of a BEC that is strongly confined in one 
dimension ($z$ direction) relative to the confinement in the other two
dimensions ($xy$ plane) can often be approximated as the product of a 
function of $x$ and $y$ only with a gaussian function of $z$ only.  In
the mean--field approximation, the actual behavior of the condensate 
wave function is governed by the 3D, time--dependent, Gross--Pitaevskii 
equation.  However, it is possible to find equations of motion 
from which the approximate product wave function can be constructed at 
each moment of time using the Lagrangian Variational Method.  We briefly 
describe this method next.

\subsection{The Lagrangian variational method}
\label{lvm}

The Lagrangian Variational Method provides approximate solutions
to the 3D TDGPE in the form of equations of motion for time--dependent
parameters that appear in an assumed trial wave function.  Thus, in 
the standard LVM, the exact solution of the 3D TDGPE requiring the
solution of a partial differential equation in three space and one time
variable is traded for the solution of ordinary differential equations
in time for the variational parameters of a trial wave function of fixed
functional form.

The 3D TDGPE can be written as
\begin{equation}
i\hbar\frac{\partial\Psi}{\partial t} = 
-\frac{\hbar^{2}}{2M}\nabla^{2}\Psi + V_{\rm trap}({\bf r})\Psi
+ gN\left|\Psi\right|^{2}\Psi,
\end{equation}
where $M$ is the mass of a condensate atom, $g=4\pi\hbar^{2}a/M$ is
the interaction strength of low--energy binary scattering events with 
$a$ being the $s$--wave scattering length, $N$ is the number of 
atoms in the condensate, and $V_{\rm trap}({\bf r})$ is the external
potential.  

The TDGPE is itself a variational equation--of--motion and is derived 
from the following Lagrangian density:
\begin{eqnarray}
{\cal L}[\Psi]
&=&
\tfrac{1}{2}i\hbar
\left(
\Psi\Psi_{t}^{\ast}-\Psi^{\ast}\Psi_{t}
\right) +
\tfrac{\hbar^{2}}{2M}
\sum_{\eta=x,y,z}
\Psi_{\eta}^{\ast}\Psi_{\eta}\nonumber\\
&+&
V_{\rm trap}({\bf r})
\Psi^{\ast}\Psi +
\tfrac{1}{2}gN
\left(\Psi^{\ast}\right)^{2}
\left(\Psi\right)^{2}
\label{tdgpe_lag_den}
\end{eqnarray}
where $\Psi_{\eta}\equiv\partial\Psi/\partial\eta$ and $\eta=x,y,z,t$.  
The associated Euler--Lagrange equation that produces the TDGPE with 
the above Lagrangian density is given by
\begin{equation}
\sum_{\eta = x,y,z,t}
\frac{\partial}{\partial\eta}
\left(
\frac{\partial\cal{L}}{\partial\Psi_{\eta}^{\ast}}
\right) -
\frac{\partial\cal{L}}{\partial\Psi^{\ast}} = 0
\label{E-L_eqn}
\end{equation}

The LVM is an approximation method that produces an equation of motion 
for the $n$ time--dependent variational parameters, $q_{k}(t), k=1,\dots,n$, 
appearing in a given trial wave function $\Psi=\psi_{\rm trial}({\bf r};
q_{1},\dots,q_{n})$.  The equations of motion for these parameters 
are obtained by inserting the trial wave function into the LVM Lagrangian 
density, integrating this over the spatial variables:
\begin{equation}
L_{\rm LVM}(q_{1}(t),\dots,q_{n}(t)) = 
\int d^{3}r\,{\cal L}\left[\psi_{\rm trial}({\bf r},t)\right],
\label{lvm_lag}
\end{equation}
and applying the usual Euler--Lagrange equations:
\begin{equation}
\frac{d}{dt}
\left(\frac{\partial L_{\rm LVM}}{\partial\dot{q}_{j}}\right) -
\frac{\partial L_{\rm LVM}}{\partial q_{j}} = 0.
\quad
j = 1,\dots,n.
\label{normal_E-L_eqs}
\end{equation}
This is the standard Lagrangian Variational Method~\cite{PhysRevA.56.1424}.  

The LVM can be regarded as having two limits in terms of the chosen trial 
wave function.  The first limit consists of choosing a trial wave function 
where the variational ``parameter'' is $\Psi({\bf r},t)$.  This choice enables 
the variational solution to vary in any possible way.  As noted above, when 
Eq.\ (\ref{E-L_eqn}) is applied to the Lagrangian density ${\cal L}[\Psi]$ 
to derive the equation of motion, it turns out to be the full TDGPE. In the 
other limit, the trial wave function is chosen to have a fixed functional 
form of the spatial coordinates where the time dependence resides entirely 
within a set of variational parameters, $\{q_{j}(t)\}_{j=1}^{n}$, so that 
the Lagrangian depends only on these parameters, $L_{\rm LVM}[q_{1},\dots,
q_{n}]$.  The shape of this trial wave function can only be varied by changing 
the values of the $q_{j}$.  The equations of motion for the $q_{j}(t)$ are 
ordinary differential equations in time and are obtained from the usual 
Euler--Lagrange equations, Eqs.\ (\ref{normal_E-L_eqs}). It is also possible 
to choose a ``hybrid'' trial wave function that plots a course midway between 
these two limits.  We describe this approach now.

\subsection{The hybrid LVM}
\label{hybrid_lvm}

The ``Hybrid Lagrangian Varational Method'' (HLVM) is an LVM in which the 
trial wave function consists of a completely unspecified function of some 
of the spatial coordinates, $\phi(x,y,t)$, multiplied by a fixed function 
of the rest of the coordinates that also contains some time--dependent
variational parameters, $q_{1}(t),\dots,q_{n}(t)$.  The HLVM is expected 
to apply to systems where there is tight confinement in one or two 
dimensions.  The coordinates appearing in $\phi$ are those for which the
confinement is weak while the trial wave function is assumed to be gaussian 
in the coordinates of tight confinment.  The HLVM for tight confinement in 
two dimensions has been studied earlier~\cite{1d_hlvm_paper}.  Here we
study the case where there is tight confinement in one dimension only.

Coupled equations of motion can be derived from a ``hybrid'' Lagrangian 
which is constructed by integrating the Lagrangian density ${\cal L}$ in
Eq.\ (\ref{tdgpe_lag_den}) over the space coordinate(s) of the tightly 
confined direction(s). The resulting hybrid Lagrangian can be used to derive 
coupled equations of motion for both $\phi$ and the set of variational 
parameters $\{q_{1},\dots,q_{n}\}$.

Before proceeding with the derivation of these equations of motion, we will
first introduce scaled variables and rewrite the LVM equations in terms of
these variables.  Scaled units are referenced to a chosen unit of length,
denoted by $L_{0}$, and scaled spatial coordinates are given by
\begin{equation}
\bar{x}\equiv\frac{x}{L_{0}},
\quad
\bar{y}\equiv\frac{y}{L_{0}},
\quad
\bar{z}\equiv\frac{x}{L_{0}}.
\label{scaled_space_vars}
\end{equation}
Energy and time units are defined in terms of $L_{0}$ enabling the definition
of a scaled time:
\begin{equation}
E_{0}\equiv\frac{\hbar^{2}}{2ML_{0}^{2}},
\quad
T_{0}\equiv\frac{\hbar}{E_{0}},
\quad
\bar{t}\equiv\frac{t}{T_{0}}.
\end{equation}
Hereafter barred symbols will denote quantities expressed in their 
appropriate scaled units.  It will also be convenient to express the 
solution of the 3D TDGPE in terms of scaled units:
\begin{equation}
\Psi\left({\bf r},t\right) \equiv L_{0}^{-3/2}
\Phi\left({\bf \bar{r}},\bar{t}\right).
\end{equation}
In terms of these variables the TDGPE becomes
\begin{equation}
i\frac{\partial\Phi}{\partial\bar{t}} = 
-\left( 
\frac{\partial^{2}\Phi}{\partial\bar{x}^{2}} +
\frac{\partial^{2}\Phi}{\partial\bar{y}^{2}} +
\frac{\partial^{2}\Phi}{\partial\bar{z}^{2}}
\right) + 
\bar{V}_{\rm trap}\Phi + 
\bar{g}N\left|\Phi\right|^{2}\Phi,
\end{equation}
where $\bar{g}\equiv g/(E_{0}L_{0}^{3})$.  In scaled units, the Lagrangian
density takes the form:
\begin{eqnarray}
\bar{\cal L}[\Phi]
&=& 
\tfrac{1}{2}i
\left(
\Phi\Phi_{\bar{t}}^{\ast} -
\Phi^{\ast}\Phi_{\bar{t}} 
\right) +
\left(
\Phi^{\ast}_{\bar{x}}\Phi_{\bar{x}} + 
\Phi^{\ast}_{\bar{y}}\Phi_{\bar{y}} +
\Phi^{\ast}_{\bar{z}}\Phi_{\bar{z}}
\right)\nonumber\\
&+&
\bar{V}_{\rm trap}\Phi^{\ast}\Phi +
\tfrac{1}{2}
\bar{g}N\left(\Phi^{\ast}\right )^{2}
\Phi^{2},
\end{eqnarray}
and the scaled Euler--Lagrange equation becomes:
\begin{equation}
\sum_{\eta = \bar{x},\bar{y},\bar{z},\bar{t}}
\frac{\partial}{\partial\eta}
\left(
\frac{\partial\cal{\bar{L}}}{\partial\Phi_{\eta}^{\ast}}
\right) -
\frac{\partial\cal{\bar{L}}}{\partial\Phi^{\ast}} = 0
\label{scaled_E-L_eqn}
\end{equation}
Now we turn to the description of the hybrid Lagrangian Variational
Method.

In deriving the HLVM equations of motion we will assume that the trapping
potential can be written (at least approximately) as the sum of a part
that depends only on the loosely confined coordinates (here $x$ and $y$)
and a part that is harmonic in the tightly bound direction.  Under this
assumption we can write the potential as:
\begin{equation}
\bar{V}_{\rm trap}(\bar{x},\bar{y},\bar{z}) \equiv
\bar{V}_{\parallel}(\bar{x},\bar{y})+\lambda^{2}\bar{z}^{2}.
\label{separable_potential}
\end{equation}
This form of the potential applies in many realistic experimental cases
such as the painted potentials mentioned earlier.

The trial wave function for the HLVM equations of motion is written as
follows:
\begin{equation}
\Phi_{\rm trial}(\bar{x},\bar{y},\bar{z},\bar{t}) = 
\phi(\bar{x},\bar{y},\bar{t})
A(\bar{t})e^{-\bar{z}^{2}/2\bar{w}^{2}(\bar{t})+
i\bar{\beta}(\bar{t})\bar{z}^{2}}.
\label{hlvm_trial_wf}
\end{equation}
Here the trial wave function is a product of a completely unspecified 
function $\phi(\bar{x},\bar{y},\bar{t})$ with a gaussian function having a 
time--dependent width $\bar{w}(\bar{t})$ and quadratic phase coefficient
$\bar{\beta}(\bar{t})$. These are the variational parameters that will 
appear in the HLVM equations of motion.  

The first step in the hybrid LVM consists of constructing a ``hybrid''
Lagrangian by integrating only over the spatial coordinate along which 
the system is strongly confined:
\begin{equation}
\bar{L}_{\rm hybrid}[\phi,A,\bar{w},\bar{\beta}]
\equiv
\int_{-\infty}^{\infty}d\bar{z}
\bar{\cal L}
\left[
\phi A
e^{-\bar{z}^{2}/2\bar{w}^{2}+i\bar{\beta}\bar{z}^{2}}
\right].
\end{equation}
The resulting hybrid Lagrangian is given by
\begin{eqnarray}
\bar{L}_{\rm hybrid}[\phi,\bar{w},\bar{\beta}]
&=&
\tfrac{i}{2}
\left(\phi\phi^{\ast}_{\bar{t}} - \phi^{\ast}\phi_{\bar{t}}\right) +
\phi_{\bar{x}}^{\ast}\phi_{\bar{x}} +
\phi_{\bar{y}}^{\ast}\phi_{\bar{y}}\nonumber\\
&+&
\phi^{\ast}\phi
\bigg(
\tfrac{1}{2}\dot{\bar{\beta}}\bar{w}^{2} + 
\tfrac{1}{2\bar{w}^{2}} + 
2\bar{\beta}^{2}\bar{w}^{2} +
\bar{V}_{\parallel}\nonumber\\
&+&
\tfrac{1}{2}\lambda^{2}\bar{w}^{2}
\bigg) +
\tfrac{1}{2}\bar{g}N
\left(\phi^{\ast}\right)^{2}\left(\phi\right)^{2}
\left(\tfrac{1}{\sqrt{2\pi}\bar{w}}\right).\nonumber\\
\end{eqnarray}
In the above we have eliminated the variational parameter $A$ using the 
normalization constraint:
\begin{equation}
\int\,d^{3}\bar{r}\left|\Phi\right|^{2} =
\left(
\int_{-\infty}^{\infty}d\bar{x}\int_{-\infty}^{\infty}d\bar{y}
\left|\phi\right|^{2}
\right)
\left(
\left|A\right|^{2}\pi^{1/2}\bar{w}
\right) = 1\nonumber\\
\end{equation}
and by requiring that the separate parts of the product wave function 
to be separately normalized to unity:
\begin{equation}
\int_{-\infty}^{\infty}d\bar{x}\int_{-\infty}^{\infty}d\bar{y}
\left|\phi\right|^{2} = 1,
\quad
\left|A\right|^{2}\pi^{1/2}\bar{w} = 1.
\end{equation}

The second step in the HLVM is to apply the Euler--Lagrange equations
of motion to $\bar{L}_{\rm hybrid}$ to obtain the equations of motion.
The equation for $\phi$ is a modified version of Eq.\ (\ref{scaled_E-L_eqn}):
\begin{equation}
\frac{\partial}{\partial\bar{x}}
\left(
\frac{\partial\bar{L}_{\rm hybrid}}{\partial\phi_{\bar{x}}^{\ast}}
\right) +
\frac{\partial}{\partial\bar{y}}
\left(
\frac{\partial\bar{L}_{\rm hybrid}}{\partial\phi_{\bar{y}}^{\ast}}
\right) -
\frac{\partial\bar{L}_{\rm hybrid}}{\partial\phi^{\ast}} = 0
\label{phi_e-l_eq}
\end{equation}
and the Euler--Lagrange equations for $\bar{w}$ and $\bar{\beta}$ are
the usual ones:
\begin{equation}
\frac{d}{d\bar{t}}
\left(
\frac{\partial\bar{L}_{\rm hybrid}}{\partial\dot{q}}
\right) - 
\frac{\partial\bar{L}_{\rm hybrid}}{\partial q} = 0,
\quad
q = \bar{w}, \bar{\beta}.
\end{equation}
Applying Eq.\ (\ref{phi_e-l_eq}) yields the following equation for
$\phi$:
\begin{eqnarray}
i\frac{\partial\phi}{\partial\bar{t}} 
&=& 
-\left(
\frac{\partial^{2}\phi}{\partial\bar{x}^{2}} +
\frac{\partial^{2}\phi}{\partial\bar{y}^{2}}
\right) + 
\bar{V}(\bar{x},\bar{y})\phi +
\left(\frac{\bar{g}N}{\sqrt{2\pi}\bar{w}}\right)
\left|\phi\right|^{2}\phi\nonumber\\
&+&
F(\bar{t})\phi
\end{eqnarray}
where
\begin{equation}
F(\bar{t}) \equiv \tfrac{1}{2}\dot{\bar{\beta}}\bar{w}^{2} +
\tfrac{1}{2\bar{w}^{2}}+2\bar{\beta}^{2}\bar{w}^{2} +
\tfrac{1}{2}\lambda^{2}\bar{w}^{2}.
\end{equation}
This seemingly complicated function of $\bar{t}$ can be transformed
away by defining
\begin{equation}
\phi(\bar{x},\bar{y},\bar{t}) \equiv
\tilde{\phi}(\bar{x},\bar{y},\bar{t})
e^{-i\int_{0}^{\bar{t}}F(\bar{t}^{\prime})d\bar{t}^{\prime}}
\label{tilde_phi_def}
\end{equation}
Inserting this into the equation of motion for $\phi$ yields an
effective 2D Gross--Pitaevskii--like equation for $\tilde{\phi}$:
\begin{eqnarray}
i\frac{\partial\tilde{\phi}}{\partial\bar{t}} 
&=& 
-\left(
\frac{\partial^{2}\tilde{\phi}}{\partial\bar{x}^{2}} +
\frac{\partial^{2}\tilde{\phi}}{\partial\bar{y}^{2}}
\right) + 
\bar{V}(\bar{x},\bar{y})\tilde{\phi} +
\left(\frac{\bar{g}N}{\sqrt{2\pi}\bar{w}}\right)
\left|\tilde{\phi}\right|^{2}\tilde{\phi}\nonumber\\
\label{phi_eom}
\end{eqnarray}

Applying the Euler--Lagrange equation for $\bar{\beta}$ gives the
following result:
\begin{eqnarray}
\left(
\frac{\partial}{\partial\bar{t}}
\left|\phi\right|^{2}
\right)
\left(
\tfrac{1}{2}\bar{w}^{2}
\right) +
\left|\phi\right|^{2}
\left(\bar{w}\dot{\bar{w}} -
4\bar{\beta}\bar{w}^{2}\right) = 0.
\end{eqnarray}
We can obtain a simplified equation of motion by integrating
both sides of the above over all $\bar{x}$ and $\bar{y}$:
\begin{eqnarray}
\left(
\tfrac{1}{2}\bar{w}^{2}
\right)
\int_{-\infty}^{\infty}d\bar{y}
\int_{-\infty}^{\infty}d\bar{x}
\frac{\partial}{\partial\bar{t}}
\left|\phi\right|^{2}
&=& 
\left(\bar{w}\dot{\bar{w}} -
4\bar{\beta}\bar{w}^{2}\right)\nonumber\\
&\times&
\int_{-\infty}^{\infty}d\bar{y}
\int_{-\infty}^{\infty}d\bar{x}
\left|\phi\right|^{2}\nonumber\\
\end{eqnarray}
It is easy to show that the integral on the left is zero by using
the equation of motion for $\phi$.  The integral on the right is unity
by normalization and so we obtain the following relationship between
$\bar{\beta}$ and $\dot{\bar{w}}$, $\bar{w}$:
\begin{equation}
\bar{\beta} = \frac{\dot{\bar{w}}}{4\bar{w}}.
\label{beta_eom}
\end{equation}
Thus, if $\bar{w}$ and $\dot{\bar{w}}$ are known, $\bar{\beta}$ is determined.

Applying the Euler--Lagrange equation for $\bar{w}$ gives
\begin{equation}
\left|\phi\right|^{2}
\left(
\dot{\bar{\beta}}\bar{w} -
\tfrac{1}{\bar{w}^{3}} +
4\bar{\beta}^{2}\bar{w} +
\lambda^{2}\bar{w}
\right) = 
\tfrac{1}{2}\bar{g}N
\left|\phi\right|^{4}
\left(
\frac{1}{\sqrt{2\pi}\bar{w}^{2}}
\right)
\end{equation}
Integrating this equation over all $(\bar{x},\bar{y})$ on both sides
as before we obtain
\begin{equation}
\dot{\bar{\beta}}\bar{w} +
4\bar{\beta}^{2}\bar{w} -
\tfrac{1}{\bar{w}^{3}} +
\lambda^{2}\bar{w} = 
\frac{\bar{g}NU_{\parallel}}{2\sqrt{2\pi}\bar{w}^{2}}
\label{1st_w_eom}
\end{equation} 
where
\begin{equation}
U_{\parallel}(\bar{t}) \equiv 
\int_{-\infty}^{\infty}d\bar{y}
\int_{-\infty}^{\infty}d\bar{x}
|\tilde{\phi}(\bar{x},\bar{y},\bar{t})|^{4}.
\end{equation}
Note that we have used Eq.\ (\ref{tilde_phi_def}) to replace $\phi$ with
$\tilde{\phi}$.

It is possible to eliminate $\bar{\beta}$ from the above equation by
differentiating both sides of Eq.\ (\ref{beta_eom}) with respect to time.
We obtain 
\begin{equation}
\tfrac{1}{4}\ddot{\bar{w}} =
\dot{\bar{\beta}}\bar{w} +
\bar{\beta}\dot{\bar{w}} = 
\dot{\bar{\beta}}\bar{w} +
4\bar{\beta}^{2}\bar{w},
\end{equation}
where the second equality results from using Eq.\ (\ref{beta_eom}) to
replace $\dot{\bar{w}}$ with $4\bar{\beta}\bar{w}$.  Now we see that
the right--hand--side of the above equation is identical to the first 
two terms on the left--hand--side of Eq.\ (\ref{1st_w_eom}). Thus we 
can rewrite this equation as follows:
\begin{equation}
\ddot{\bar{w}} +
4\lambda^{2}\bar{w} = 
\frac{4}{\bar{w}^{3}} +
\frac{\sqrt{2/\pi}\bar{g}NU_{\parallel}}{\bar{w}^{2}}
\label{w_eom}
\end{equation}
This is the final equation of motion for $\bar{w}$.

\begin{figure*}[htb]
\begin{center}
\mbox{\psboxto(7.0in;0.0in){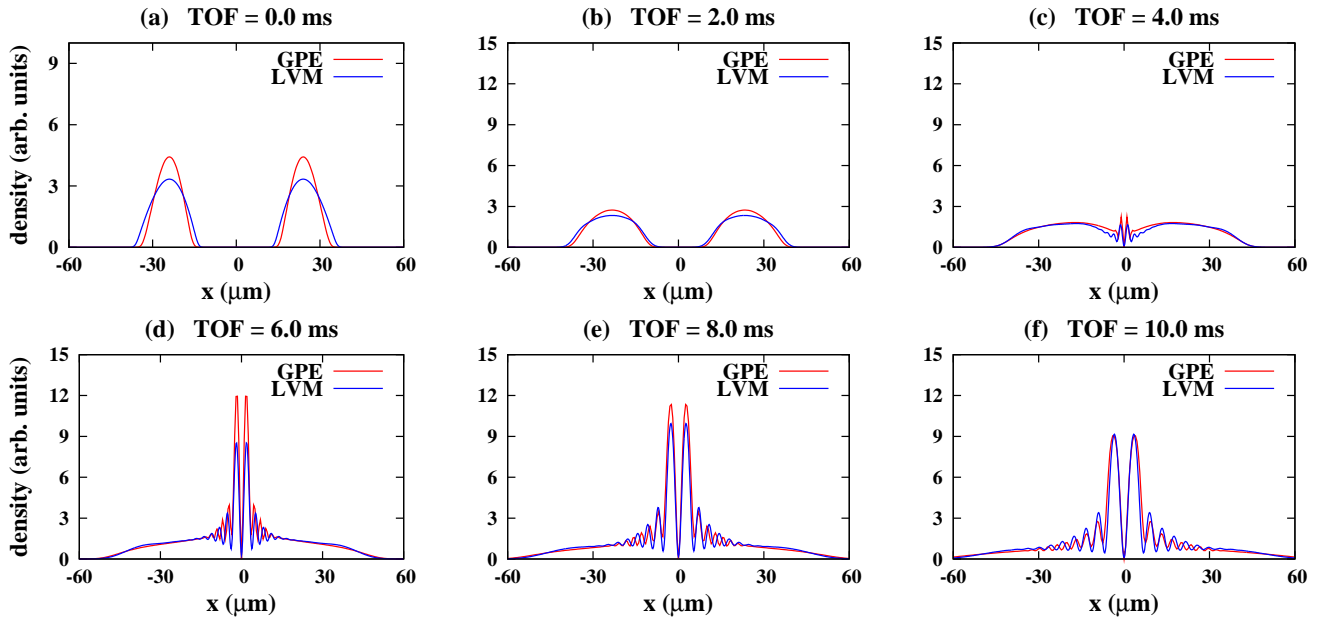}}
\caption{(color online) A comparison of the vertical column density of a 
Bose--Einstein condensate, as determined by the 3D TDGPE and the HLVM equations
of motion, after direct release from a ring--shaped trap is plotted along a line 
through the trap center ($x$ axis) for various times of flight (TOF) during 
expansion, (a) TOF = 0.0 ms, (b) TOF = 2.0 ms, (c) TOF = 4.0 ms, (d) TOF = 6.0 
ms, (e) TOF = 8.0 ms, and (f) TOF = 10.0 ms.  The condensate is given one 
unit of angular momentum before release by phase imprint.}
\label{direct_time_compare_fig}
\end{center}
\label{time_cmp_fig}
\end{figure*} 

\subsection{The HLVM equations of motion and the variational 
initial state}
\label{hlvm_eom}

The full set of HLVM equations of motion consist of a 2D effective
GP--like equation for $\tilde{\phi}$:
\begin{eqnarray}
i\frac{\partial\tilde{\phi}}{\partial\bar{t}} 
&=& 
-\left(
\frac{\partial^{2}\tilde{\phi}}{\partial\bar{x}^{2}} +
\frac{\partial^{2}\tilde{\phi}}{\partial\bar{y}^{2}}
\right) + 
\bar{V}(\bar{x},\bar{y})\tilde{\phi} +
\left(\frac{\bar{g}N}{\sqrt{2\pi}\bar{w}}\right)
\left|\tilde{\phi}\right|^{2}\tilde{\phi}\nonumber\\
\label{hlvm_phi_eq}
\end{eqnarray}
and an equation for $\bar{w}$:
\begin{equation}
\ddot{\bar{w}} +
4\lambda^{2}\bar{w} = 
\frac{4}{\bar{w}^{3}} +
\frac{\sqrt{2/\pi}\bar{g}NU_{\parallel}}{\bar{w}^{2}}.
\label{hlvm_w_eq}
\end{equation}
These two equations form a closed system from which $\bar{w}(\bar{t})$,
$\dot{\bar{w}}(\bar{t})$, and $\tilde{\phi}(\bar{x},\bar{y},\bar{t})$
can be obtained.  From these, the value of $\bar{\beta}(\bar{t})$ and 
$F(\bar{t})$ can be calculated:
\begin{equation}
\bar{\beta} = \frac{\dot{\bar{w}}}{4\bar{w}}
\quad
F(\bar{t}) = \tfrac{1}{2}\dot{\bar{\beta}}\bar{w}^{2} +
\tfrac{1}{2\bar{w}^{2}}+2\bar{\beta}^{2}\bar{w}^{2} +
\tfrac{1}{2}\lambda^{2}\bar{w}^{2}.
\label{beta_and_F}
\end{equation}
Using these quantities, the full value of the variational trial wave
function can be constructed:
\begin{eqnarray}
\Phi_{\rm trial}(\bar{x},\bar{y},\bar{z},\bar{t}) 
&=& 
\left(
\frac{1}{\pi^{1/2}\bar{w}}
\right)^{1/2}
\tilde{\phi}(\bar{x},\bar{y},\bar{t})
e^{-i\int_{0}^{\bar{t}}F(\bar{t}^{\prime})d\bar{t}^{\prime}}\nonumber\\
&\times&
e^{-\bar{z}^{2}/2\bar{w}^{2}(\bar{t})+i\bar{\beta}(\bar{t})\bar{z}^{2}},
\label{hlvm_wf}
\end{eqnarray}
where
\begin{equation}
U_{\parallel}(\bar{t}) \equiv 
\int_{-\infty}^{\infty}d\bar{y}
\int_{-\infty}^{\infty}d\bar{x}
|\tilde{\phi}(\bar{x},\bar{y},\bar{t})|^{4}.
\label{U_parallel}
\end{equation}
Note that Eqs.\ (\ref{hlvm_phi_eq}) and (\ref{hlvm_w_eq}) are coupled.  
The nonlinear term in the 2D GPE for $\tilde{\phi}$ contains the gaussian 
width, $\bar{w}$, while the equation for $\bar{w}$ contains the factor
$U_{\parallel}$ which is the integral of the fourth power of  $\tilde{\phi}$.

The final element required for this method to be used as a means to 
find an approximation to the solution of the 3D TDGPE is a set of initial 
conditions.  We present one possibility here based on the physics of
Bose--Einstein condensate systems.

In a typical BEC experiment a condensate is formed in an atom trap.  If
no further changes in the condensate's environment occur, the condensate
wave function should then, in principle, only acquire an overall time--%
dependent phase as it evolves in time.  In the HLVM this situation should 
therefore be represented by the {\em stationary} solution of the above 
equations.  We denote this stationary solution as
\begin{equation}
\tilde{\phi}(\bar{x},\bar{y},0) \equiv 
\tilde{\phi}_{0}(\bar{x},\bar{y})
\quad{\rm and}\quad
\bar{w}(0) \equiv w_{0}.
\end{equation}
This solution satifies the following time--independent equations.
\begin{eqnarray}
\mu\tilde{\phi}_{0}
&=& 
-\left(
\frac{\partial^{2}\tilde{\phi}_{0}}{\partial\bar{x}^{2}} +
\frac{\partial^{2}\tilde{\phi}_{0}}{\partial\bar{y}^{2}}
\right) + 
\bar{V}(\bar{x},\bar{y})\tilde{\phi}_{0}\nonumber\\ 
&+&
\left(\frac{\bar{g}N}{\sqrt{2\pi}\bar{w}_{0}}\right)
\left|\tilde{\phi}_{0}\right|^{2}\tilde{\phi}_{0}
\label{initial_phi_eq}
\end{eqnarray}
and
\begin{equation}
R(\bar{w}_{0}) \equiv
4\lambda^{2}\bar{w}_{0} - 
\frac{4}{\bar{w}_{0}^{3}} -
\frac{\sqrt{2/\pi}\bar{g}NU_{\parallel,0}}{\bar{w}_{0}^{2}} = 0.
\label{initial_w_eq}
\end{equation}
where
\begin{equation}
U_{\parallel,0} \equiv 
\int_{-\infty}^{\infty}d\bar{y}
\int_{-\infty}^{\infty}d\bar{x}
|\tilde{\phi}_{0}(\bar{x},\bar{y})|^{4}.
\end{equation}
The factor $\mu$ in the equation for $\tilde{\phi}$ is the chemical
potential of the initial condensate. These equations for the stationary
variational solution must be solved self--consistently.

\begin{figure*}[htb]
\begin{center}
\mbox{\psboxto(7.0in;0.0in){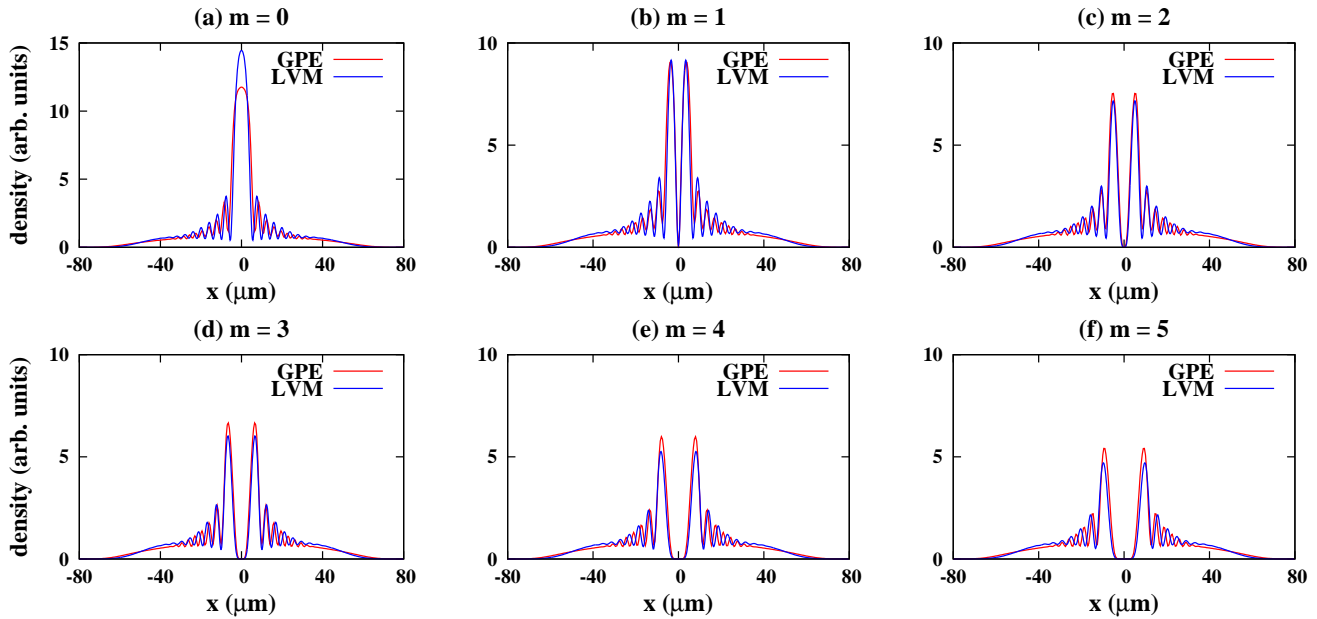}}
\end{center}
\caption{(color online) GPE/LVM vertical column density comparison for different
angular momenta applied to the initial state.  The condensates are formed,
stirred to add $m$ units of angular momentum (simulated by phase imprint), and
then released and allowed to expand for 10 ms. (a) $m = 0$, (b) $m = 1$, (c) 
$m = 2$, (d) $m = 3$, (e) $m = 4$, and (f) $m = 5$.}
\label{winding_compare_fig}
\end{figure*} 

\section{Comparison with 3D GPE}
\label{2d3d_compare}

In this section we illustrate the ability of the HLVM equations of motion 
to approximate the exact solution of the 3D TDGPE by comparing the two 
solutions for a case of current interest.  The system we will consider is 
that of a Bose--Einstein condensate of $^{23}$Na atoms confined in a ring--%
shaped potential under the same conditions as in a recent experiment~\cite
{PhysRevLett.106.130401} conducted at NIST.  To simplify the analysis we 
will only compute the profile of the condensate density integrated along
the vertical direction for each point in the plane of weak confinement. 
This quantity predicted by the TDGPE will be compared with that predicted by 
the HLVM equations of motion.  This is the quantity that can be compared with
experiment. 

In the NIST experiment a vertical Laguerre--Gauss (LG) laser beam (LG$_{0}^{1}$) 
was intersected with a horizontal light sheet.  The shape of the vertical
LG beam was approximately a hollow cylinder with thick walls so that its
intersection with the horizontal light sheet created a ring--shaped region
of maximum light intensity.  Tuning the frequency of the beams to the red
of the lowest electronic transition created a potential that caused the atoms
to seek the maximum intensity.  

In this comparison, we simulate an experiment in which a condensate is created 
in this ring potential, optionally stirred, and then probed.  We simulate two 
types of probes: (1) direct release of the condensate by turning off all trapping 
potentials after stirring, and (2) release of the condensate after the ramp down
of the Laguerre--Gauss potential.  The stirring, which adds $m$ units of 
angular momentum to the condensate, is simulated by phase imprint.  That is, the 
initial condensate wave function is multiplied by $e^{im\phi}$ where $\phi$ is 
the azimuthal angle around the vertical $\bar{z}$ axis and $m$ is an integer.

In each case we will compare what would be the measured density profile, as 
predicted by the 3D GPE and by the HLVM, for different times during the ramp 
down or expansion where the value of $m$ is fixed, and also for a fixed final 
time--of--flight for a range of different $m$ values.  For maximum clarity, 
we present the two density profiles as a plot of the density along a line that 
cuts through the center of the ring.  Since all of the density profiles are 
cylindrically symmetric, these plots will convey all of the available density
information.

In these simulations the trap potential is modeled as the sum of a Laguerre--%
Gauss optical potential~\cite{PhysRevA.63.013608} plus a vertical gaussian due 
to the light sheet.  This potential can be written (in scaled units) as:
\begin{eqnarray}
\bar{V}_{\rm trap}(\bar{x},\bar{y},\bar{z},\bar{t}) 
&=& 
-e\bar{V}_{LG}f(\bar{t})
\left(
\frac{\bar{x}^{2}+\bar{y}^{2}}{\bar{r}_{M}^{2}}
\right)
e^{(\bar{x}^{2}+\bar{y}^{2})/\bar{r}_{M}^{2}}\nonumber\\
&-&
\bar{V}_{sheet} +
\lambda^{2}\bar{z}^{2} \equiv
\bar{U}_{\parallel}(\bar{x},\bar{y}) + 
\lambda^{2}\bar{z}^{2},\nonumber\\
\end{eqnarray}
where the factor $e=2.718\dots$ is included so that $\bar{V}_{LG}$ becomes the 
depth of the potential due to the Laguerre--Gaussian beam and $\bar{r}_{M}$ is 
the radial position of its minimum.  A time--dependent, dimensionless turn--on 
function, $0\le f(\bar{t})\le 1$, is inserted to simulate the ramp down of the 
Laguerre--Gauss potential. The factor $\bar{V}_{sheet}$ is the depth of the 
potential due to the light sheet and the $\bar{z}$--dependent gaussian light%
--sheet potential has been approximated by an harmonic oscillator.

\begin{figure*}[htb]
\begin{center}
\mbox{\psboxto(7.0in;0.0in){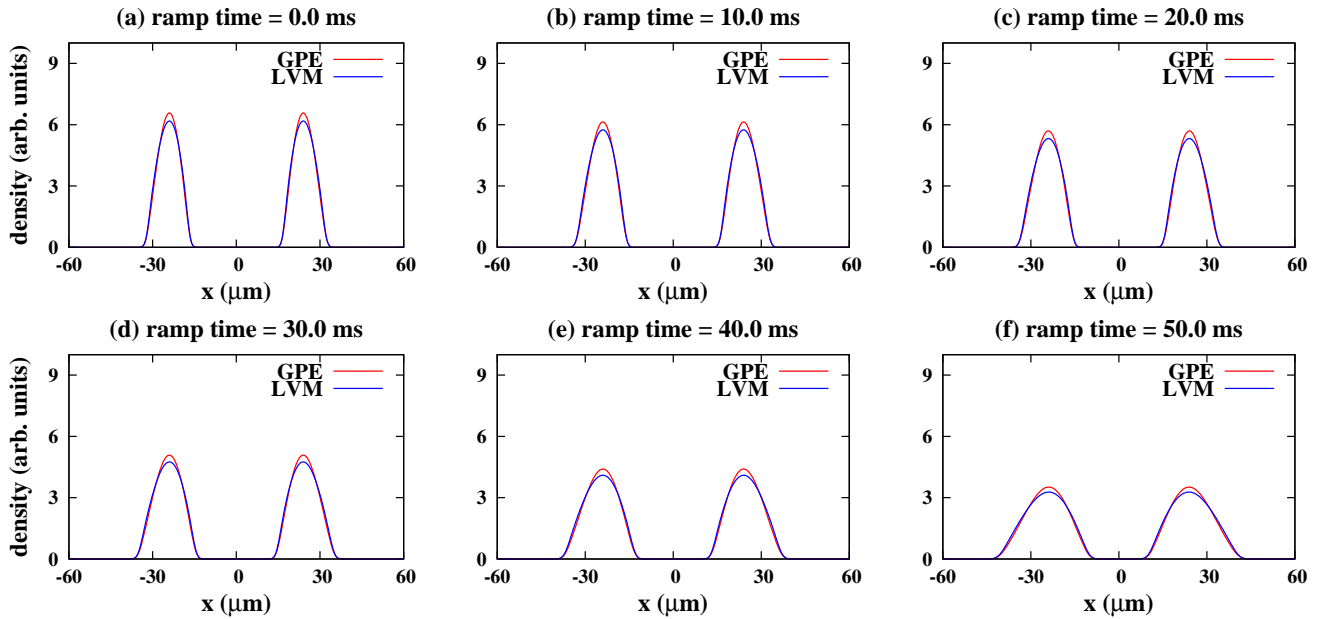}}
\end{center}
\caption{(color online) GPE/LVM vertical column density comparison for the
case where the LG potential ramped down to 20\% of its initial depth over
a span of 50 ms.  A phase imprint is applied that simulates one unit of
angular momentum added by stirring.  The plots show the comparison for ramp
times of (a) 0 ms, (b) 10 ms, (c) 20 ms, (d) 30 ms, (e) 40 ms, and (f) 
50 ms.}
\label{rampdown_compare_fig}
\end{figure*} 

Both the 3D TDGPE and the 2D GPE part of the HLVM equations of motion were solved 
using the split--step, Crank-Nicolson method.  The 3D TDGPE was solved on a 
grid in which there were 400 points along $x$ and $y$ and 200 points along $z$.
The 2D GPE part of the HLVM equations of motion was solved on a grid of 800
points along both $x$ and $y$.  The codes that were used to solve these equations 
were extensively modified versions of codes publicly available in the literature%
~\cite{Muruganandam20091888}.  The initial condensate wave function for the 3D 
TDGPE was obtained by solving it in imaginary time.  Initial conditions for the 
HLVM equation of motion were obtained by solving equations (\ref{initial_phi_eq}) 
and (\ref{initial_w_eq}) self consistently as follows.  First, a value for 
$\bar{w}_{0}$ was chosen, the associated $\tilde{\phi}$ was then found by integrating 
Eq.\ (\ref{hlvm_phi_eq}) in imaginary time, next the value of $\bar{U}_{\parallel}$ 
was calculated which was then used to compute the value of $R(\bar{w}_{0})$ in 
Eq.\ (\ref{initial_w_eq}).  The value of $\bar{w}_{0}$ was incremented and the 
process was repeated to compute a new value of $R$.  This process was continued 
until a root of $R(\bar{w}_{0})$ was found.  The value of $\bar{w}_{0}$ for this 
case is the self--consistent gaussian width of the stationary solution of the 
HLVM equations of motion.

In the direct--release process simulated, the number of condensate atoms was
$N= $750,000 atoms and the scattering length of $^{23}$Na was taken to be 53 bohr.
The minimum of the LG potential was set at $r_{M}=24 \mu$m.  The depth of the LG
potential was taken to be $V_{LG}=227$ nK which is equivalent (via $V_{LG}=\tfrac
{1}{4}M\omega_ {r}^{2}r_{M}^{2}$) to a radial harmonic frequency of $\omega_{r}/
(2\pi) = 120$ Hz.  The frequency of the harmonic light sheet potential was taken 
as $\omega_{z}/(2\pi)=320$ Hz and the light--sheet depth was $V_{sheet} = 473$ nK 
although this last quantity makes no difference in the shape of the initial--state
density.

Figure \ref{direct_time_compare_fig} displays a comparison of the integrated
column density of a released ring BEC predicted by the 3D TDGPE with that
predicted by the HLVM equations of motion at six different times--of--flight 
(TOF) after release beginning with Fig.\ (\ref{direct_time_compare_fig}a) 
showing the moment of release.  The condensate has been stirred so that it is 
released having one unit of angular momentum.  We note that there is good
agreement with quantitative differences occuring in the heights of individual 
peaks and in the position of the peaks at later times.  The comparisons are 
typical of a variational solution in that they are the ``best fit'' to the 
exact solution for the given trial wave function.  

A comparison of the GPE and LVM results for a ring BEC directly released from 
the trap and allowed to expand for a fixed TOF for different initial angular 
momenta is exhibited in Fig.\ \ref{winding_compare_fig}.  The figure displays
comparisons for $m$ values ranging from 0 to 5.  Again the agreement is good 
although there are some quantitative differences as to positions of the individual
peaks.  It is clear that there is qualitative and almost--quantitative agreement
between the 3D TDGPE and the HLVM equations of motion for these cases.

We next compare the results of the TDGPE and HLVM for ring--BEC evolution
while the LG potential is ramped down from its initial value.  This differs 
from the previous comparison in that the confining light--sheet potential
remains unchanged during the ramp down.  In the simulated ramp--down process
the number of condensate atoms was $N= $500,000 atoms.  The LG potential 
depth was ramped linearly down from its initial value of $V_{LG}=227$ nK 
(the same as previously) to 20\% of this value over a timespan of 50 ms.  
The light sheet potential was the same as in the direct--release simulations.
A phase imprint was applied to the condensate to simulate one unit of angular
momentum added by stirring.  

Figure~\ref{rampdown_compare_fig} shows the comparison starting at $t = 0$ 
ms and for every 10 ms thereafter until the rampdown is complete.   The two 
solutions again exhibit the type of agreement that is usual for variational
approximations in that the variational solution is a ``best fit'' to the 
numerical solultion.  With that caveat, the agreement here is quite good
during the entire ramp down.

\section{Summary}
\label{summary}

In this paper we derived equations of motion whose solution approximates
the solution of the 3D TDGPE applied to a quasi--2D Bose--Einstein condensate.  
The equations were derived using a hybrid Lagrangian Variational Method.  
Similar equations of motion were derived earlier~\cite{1d_hlvm_paper} for 
quasi--1D BEC systems.  The main advantage of solving these equations is that 
numerical solution of the HLVM equation can be performed 100 to 1000 times 
faster than solving the 3D TDGPE.  In the comparison simulation presented in
Section \ref{2d3d_compare}, solving the realtime 3D TDGPE required more
than 24 hours of CPU time while solving the HLVM equation took about 10 
minutes on a commodity desktop PC.  The resulting speedup here is roughly a
factor of 150.

This advantage enables rapid simulation of many different possible quasi--2D
systems.  It should be noted that when the HLVM was applied to a quasi--1D system 
describing soliton splitting in optical fibers~\cite{trippenbach_hlvm_paper},
it was found that occasionally the HLVM equations of motion did not provide 
any advantage over the regular LVM technique.  In the work cited, the authors
recommended that any important results that come out of HLVM simulations be
confirmed by simulations using the full equations.  We did not find any case 
where the HLVM equations predicted behavior that was qualitatively different
from the 3D TDGPE.  However, we agree with the recommendation of the authors of 
Ref.\ \cite{trippenbach_hlvm_paper}.

This caveat notwithstanding, the HLVM equations derived in this paper
enable rapid study of different systems of current experimental interest.  In
particular, they should be useful in simulating time--dependent behavior of
quasi--2D atomtronic systems where mean--field theory applies.  We expect 
this approximation to become a useful tool in studying future quasi--2D
Bose--Einstein condensate systems.

\begin{acknowledgments}
This material is based upon work supported by the U.S.\ National Science
Foundation under grant numbers PHY--1004975, PHY--0758111, the Physics 
Frontier Center grant PHY--0822671 and by the National Institute of Standards
and Technology.  The authors acknowledge helpful discussions from Kevin Wright,
Gretchen Campbell, and Noel Murray.
\end{acknowledgments}

\bibliography{2d_lvm_hybrid_paper}{}

\end{document}